\documentclass{moriond}

\usepackage[numbers]{natbib}

\def\be{\begin{equation}}
\def\ee{\end{equation}}
\def\bea{\begin{eqnarray}}
\def\eea{\end{eqnarray}}

\newcommand{\Photo}{}

\begin{document}
\vspace*{4cm}
\title{THE HOMOGENEITY SCALE OF THE UNIVERSE}

\author{ P. NTELIS }

\address{Astroparticle and Cosmology Group, University Paris-Diderot 7 , 10 rue A. Domon $\&$ Duquet, Paris , France}

\maketitle
\abstracts{
In this study, we probe the cosmic homogeneity with the BOSS CMASS galaxy sample in the redshift region of $0.43 < z < 0.7$. We use the normalised counts-in-spheres estimator $\mathcal{N}(<r)$ and the fractal correlation dimension $\mathcal{D}_{2}(r)$ to assess the homogeneity scale of the universe. We verify that the universe becomes homogenous on scales greater than $\mathcal{R}_{H} \simeq 64.3\pm1.6\ h^{-1}Mpc$, consolidating the Cosmological Principle with a consistency test of $\Lambda$CDM model at the percentage level. Finally, we explore the evolution of the homogeneity scale in redshift. }

\section{Introduction on Cosmic Homogeneity}

\paragraph{}The standard model of Cosmology, known as the $\Lambda$CDM model is based on solutions of the equations of General Relativity for isotropic and homogeneous universes where the matter is mainly composed of Cold Dark Matter (CDM) and the $\Lambda$ corresponds to a cosmological constant. This model shows excellent agreement with current data, be it from Type Ia supernovae, temperature and polarisation anisotropies in the Cosmic Microwave Background or Large Scale Structure. 

The main assumption of these models is the Cosmological Principle which states that the universe is homogeneous and isotropic on large scales ~\cite{CP}. Isotropy is well tested through various probes in different redshifts such as the Cosmic Microwave Background temperature anisotropies at $z\approx1100$, corresponding to density fluctuations of the order $10^{-5}$~\cite{CMB-Cobe}. In later cosmic epochs, the distribution of source in surveys  the hypothesis of isotropy is strongly supported by in X-ray~\cite{X-ray-ref1} and radio~\cite{Radio}, while the large spectroscopic galaxy surveys such show no evidence for anisotropies in volumes of a few $\mathrm{Gpc}^3$~\cite{boss2011}.As a result, it is strongly motivated to probe the homogeneity property of our universe.

	In this study, we use a fiducial flat $\Lambda$CDM cosmological model using the parameters estimated by Planck ~\cite{Params-Planck}:
	\be\label{fid-cosmo}
		p_{cosmo}=(\omega_{cdm},\omega_{b},h,n_{s},ln\left[10^{10}A_{s}\right]) = (0.1198,0.02225,0.6727,0.9645,3.094)
	\ee

\section{Methodology}
\subsection{Dataset Characteristics}\label{subsec:DataChar}
The Baryon Oscillation Spectroscopic Survey (BOSS) is a part of Sloan Digital Sky Survey SDSS-III $2.5$m telescope located at Apache point Observatory \cite{Telescope}. BOSS is dedicated to study the clustering of $\sim1.4\times10^{6}$ galaxies and $\sim10^{5}$ of Quasars and their Lyman-a forests in an effective area of $\sim10000 deg^{2}$ \cite{BOSS-DR12}. Using $1000$ optical fibers on the focal plane of the telescope the light of each target is led to the spectrograph to allow gather the spectra of $1000$ objects per exposure~\cite{spectra}.  
Using the imaging data in u,g,r,i,z bands, massive galaxy targets are selected~\cite{Target-Selection} through several colour cuts designed to keep targets in redshifts between $0.43<z<0.7$ and keeping a constant stellar mass according to a passively evolving model ~\cite{Maraston}. We divide our sample into North and South galactic Cap and in we show the amount of objects in $5$ redshift bins as shown in the left panel of figure [\ref{fig:zobjects_bias}]. There we show for comparison the amount of objects observed on a previous study ~\cite{WiggleZ}. 

We use an FRW-metric to infer the comoving distances of galaxies from the measured spectra:
\be
	\chi_{comov} (z) = c \int_{0}^{z} \frac{dz'}{H(z')}
\ee
where $H(z) = H_{0} \sqrt{ \Omega_{m,0}(1+z)^{3} + \Omega_{\Lambda,0} }$ and $H_{0} = 100h\ \mathrm{km/s/Mpc}$.

\subsection{Estimators}
In order to determine the homogeneity scale we use the estimator for the galaxy density of our BOSS CMASS sample, the \textit{Normalised Count-in-Spheres} which is defined as follows:
\be
	\mathcal{N}(<r) = \frac{N(<r)}{N_{Random}(<r)} 
\ee
where we have divided by the count-in-spheres of a random distribution ($N_{Random}(<r)$) placed on the same galaxy sample volume in order to correct for geometric effects. A random distribution will have count-in-spheres scaling with the volume ($N_{Random}(<r) \propto r^3$) while for a fractal distribution will scaling with a power-law as  ($N(<r) \propto r^{D_{2}}$). So our main estimator is the \textit{Fractal Correlation Dimension}:
\be \label{eq:estimator}
	\mathcal{D}_{2}(r) = \frac{d\ ln\ \mathcal{N}(<r)}{ d\ ln\ r} + 3
\ee
Since this estimator is defined with the total matter distribution, e.i. the biased galaxy sample, we need to convert it using a Redshift Space Distortion Analysis as explained in section \ref{RSD-Analysis}. The new estimator is corrected for the bias of the galaxy clustering $b$ w.r.t. Total Matter and the peculiar velocities $\sigma_p$:
\be 
	\mathcal{D}_{2}(r;b,\sigma_p) = \frac{d\ ln\ \mathcal{N}(<r;b,\sigma_p)}{ d\ ln\ r} + 3
\ee
Then we chose to define the Homogeneity Scale as the scale at which the fractal correlation dimension gets the value for a homogeneous distribution at $1\%$, following WiggleZ ~\cite{WiggleZ}:
\be\label{eq:Rh}
	\mathcal{D}_{2}(\mathcal{R}_{H};b,\sigma_p) = 2.97
\ee

\section{RSD Analysis}\label{RSD-Analysis}
\subsection{Modelling}
\paragraph{}We need to reconstruct our estimator \textit{fractal correlation dimension}. To do that we use a simplistic Redshift Space Distortion Model which is the convolution of a factor $F(r;b,\sigma_{p})$ with the $2$-point-correlation-function ($2$ptCF) of total matter in our universe:
\be
	\xi(r;b,\sigma_p,p_{cosmo}) = F(r;b,\sigma_p) \otimes \xi(r;p_{cosmo})
\ee
This model assumes a linear evolution of the power spectrum which is Fourier Transform gives the 2ptCF in the real space $\xi(r;p_{cosmo})$ .To account for non linearities on small scale we include the $F(r;b,\sigma_p)$ factor. This accounts for the Kaiser effect in scales greater that $10h^{-1}Mpc$ and the Finger-of-god effect in scales less than $10h^{-1}Mpc$. 

These models assume a non-scale dependent galaxy bias at the corresponding redshifts. We use the CLASS software to obtain the linear power spectrum using our fiducial parameters $p_{cosmo}$ ~\cite{CLASS}.

\subsection{Measurement of RSD parameters}
\begin{figure}
\begin{minipage}{0.33\linewidth}
\centerline{\includegraphics[width=1.0\linewidth,draft=False]{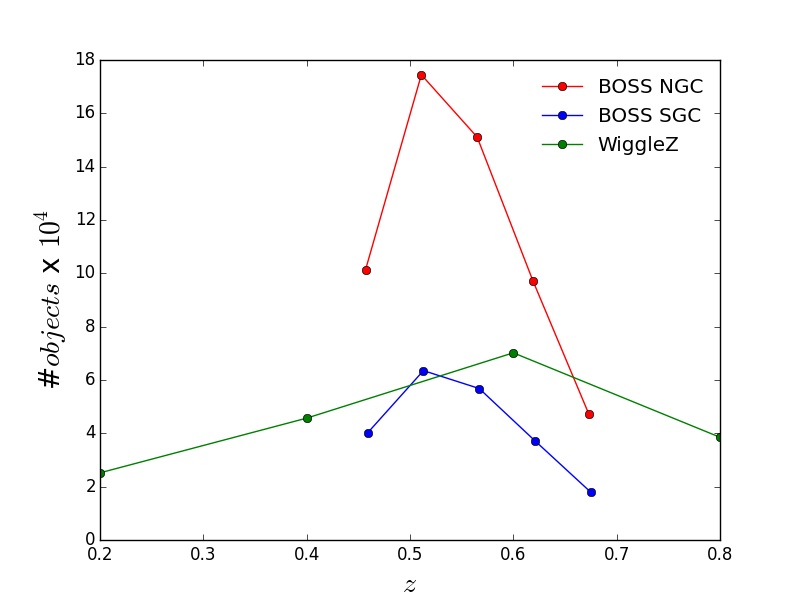}}
\end{minipage}
\hfill
\begin{minipage}{0.32\linewidth}
\centerline{\includegraphics[width=1.0\linewidth]{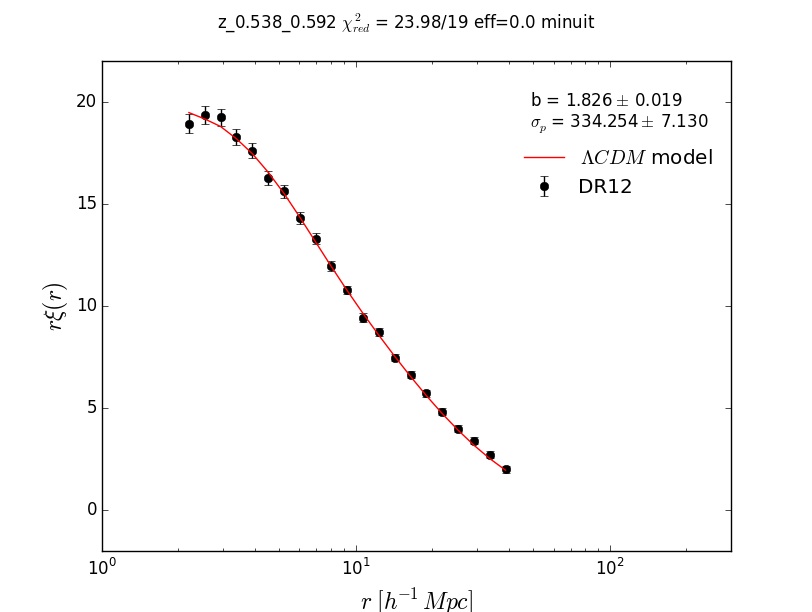}}
\end{minipage}
\hfill
\begin{minipage}{0.32\linewidth}
\centerline{\includegraphics[width=1.0\linewidth]{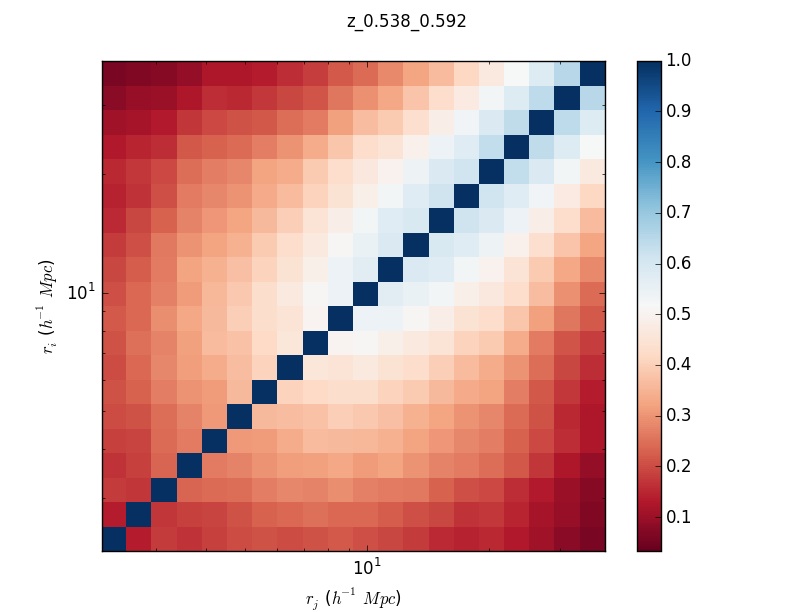}}
\end{minipage}
\caption[]{\textit{Left}: Redshift profile of objects. In red(blue) the NGC(SGC) CMASS sample. In green the WiggleZ galaxy sample ~\cite{WiggleZ}. The amount of objects of BOSS survey are 4 times more that the WiggleZ survey. \textit{Center}: RSD analysis determining the galaxy bias and peculiar velocities in the z=0.538-0.592 bin in the NGC. Black error bars correspond to the scaled 2ptCF as a function of scale. With red-line is the best-fit $\Lambda$CDM model prediction for the fiducial cosmological parameters with $1\%$ and $2\%$ measurement of galaxy bias and peculiar velocities respectively.  \textit{Right:} The corresponding correlation matrix of the 2ptCF built by $1000$ QPM mock catalogues.  }
\label{fig:zobjects_bias}
\end{figure}

\paragraph{} We tune our fitting parameters and ranges by using $1000$ QPM mocks. These mocks are constructed with quick-particle-mesh algorithm which is  faster but almost the same amount accurate as heavy N-body simulations ~\cite{QPM}. The result is shown in figure [\ref{fig:zobjects_bias}] in the middle panel. In the right panel of figure [\ref{fig:zobjects_bias}], we plot the correlation matrix of the 2ptCF. We obtain a measured galaxy bias $b=1.826\pm 0.019$ and a galaxy peculiar velocity parameter of $\sigma_p \simeq 334\pm 7 \ Km.s^{-1}$.  

We use these parameters to convert our galaxy \textit{Fractal Correlation Dimension} observable into the one corresponding for the total matter $\mathcal{D}_{2}^{MM}(r;b,\sigma_p) \rightarrow \mathcal{D}_{2}^{MM}(r)$ as it is demonstrated in the left panel of figure \ref{fig:Results}. The conversion of equation \ref{eq:estimator} is explained analytically in ~\cite{PNtelis} and references therein.

\section{Results with the Homogeneity Scale $\mathcal{R}_H$}
\subsection{Fractal Correlation Dimension}
In figure [\ref{fig:Results}] in the middle panel, the scale dependence of the \textit{fractal correlation dimension} is shown. We show that $\mathcal{D}_{2}(r)$ is increasing with scale and reaches asymptotically the nominal value $3$ which corresponds to a homogeneous universe. This provides us with the information that the universe becomes homogeneous by increasing scale which validates the Cosmological Principle in a consistent way. Furthermore, it is shown that for smaller scales the $\mathcal{D}_{2}(r)$ deviates from the nominal value implying that the universe presents a clustering or a fractal-like behaviour. 

Moreover, we measure the homogeneity scale according to equation \ref{eq:Rh} by fitting a $5$-node spline around the nominal value 2.97. We end up with $\mathcal{R}_{H} = 64.3\pm1.6h^{-1}Mpc$ within $2\sigma$ from $\Lambda$CDM prediction $\mathcal{R}^{th}_{H}=62h^{-1}Mpc$. We also show the performance of 1000 QPM catalogues which gives as the estimation of the covariance matrix needed to our calculation.

\subsection{Epoch evolution of Cosmic Homogeneity}
\begin{figure}
\begin{minipage}{0.33\linewidth}
\centerline{\includegraphics[width=1.0\linewidth,draft=False]{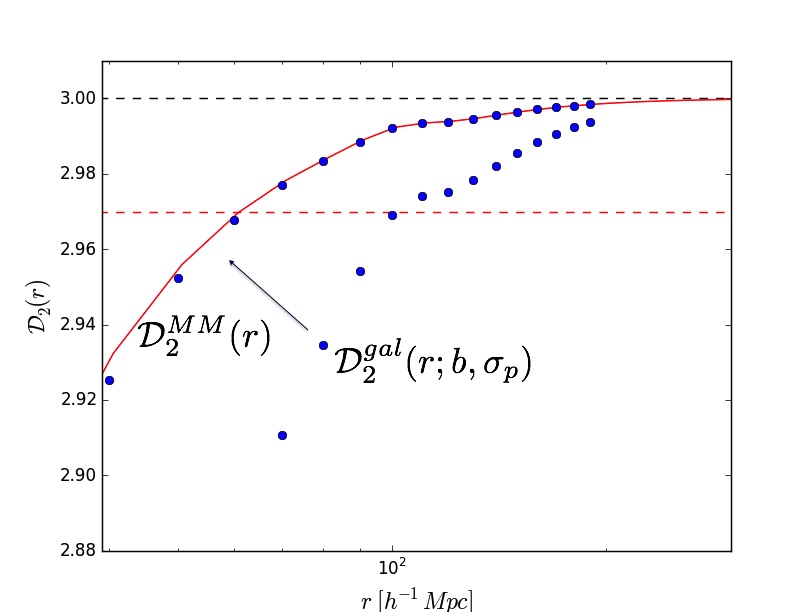}}
\end{minipage}
\hfill
\begin{minipage}{0.32\linewidth}
\centerline{\includegraphics[width=1.0\linewidth]{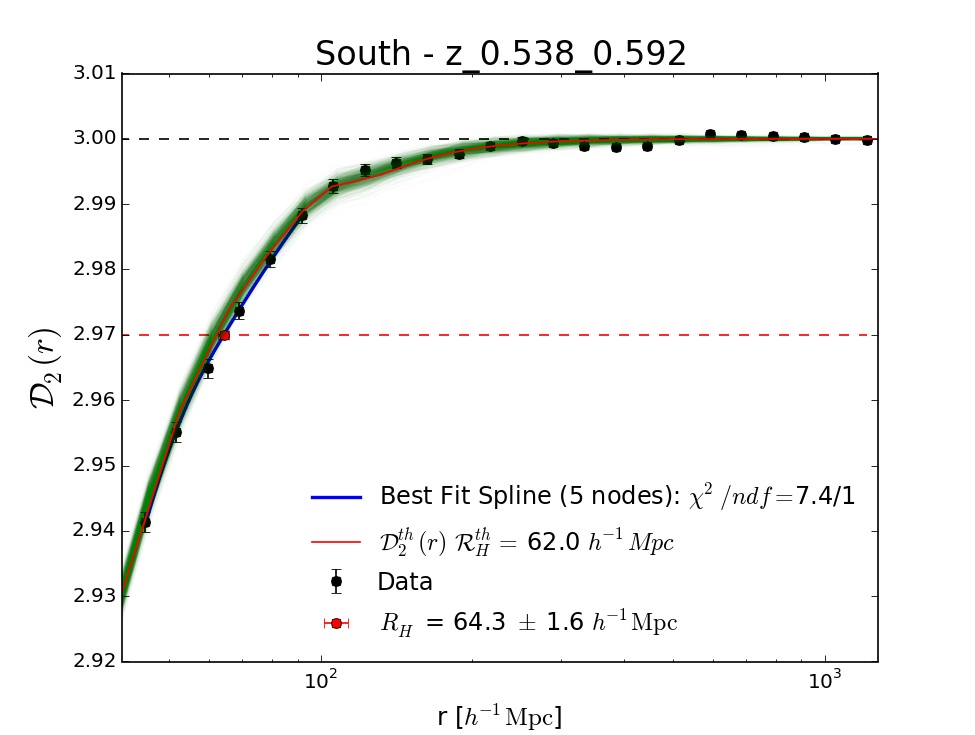}}
\end{minipage}
\hfill
\begin{minipage}{0.32\linewidth}
\centerline{\includegraphics[width=1.0\linewidth]{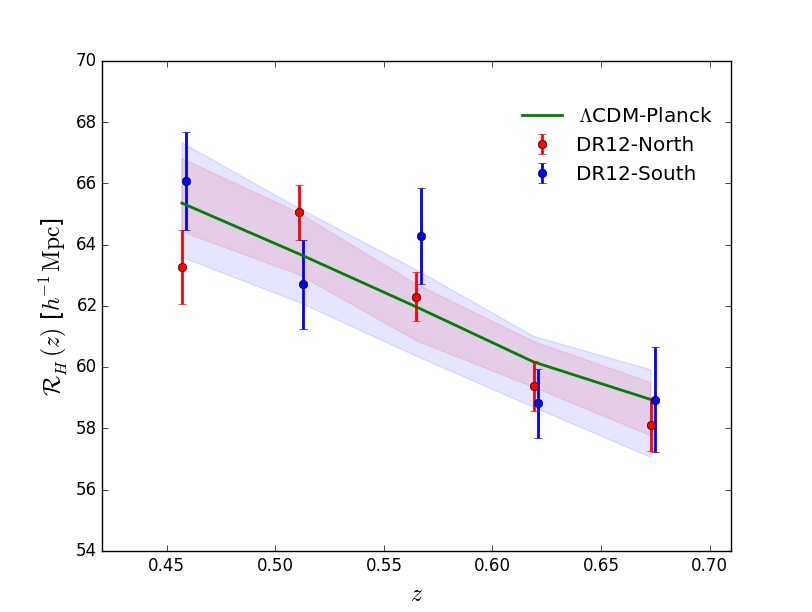}}
\end{minipage}
\caption[]{\textit{Left}:  $\mathcal{D}_{2}^{MM}(r;b,\sigma_p) \rightarrow \mathcal{D}_{2}^{MM}(r)$ conversion in order to account for the bias and redshift distortion of our cosmic matter tracer. \textit{Center}: \textit{Fractal correlation dimension} scale dependence measurement. By fitting a $5$-node spline around the nominal value 2.97 according to equation \ref{eq:Rh}, we measure the homogeneity scale $\mathcal{R}_{H} = 64.3\pm1.6h^{-1}Mpc$ at the middle z-bin of our galaxy sample. Green-lines correspond to $1000$ QPM catalogues.  \textit{Right:} Cosmic Homogeneity redshift evolution. Red(Blue) corresponds to NGC(SGC) CMASS galaxy sample in accordance with the $\Lambda$CDM prediction. We can infer the growth of structures with time. }
\label{fig:Results}
\end{figure}

\paragraph{}In figure [\ref{fig:Results}] in the middle panel, we plot the epoch dependence of the homogeneity scale $\mathcal{R}_{H}=\mathcal{R}_{H}(z)$ at $0.43<z<7$. We show that the homogeneity scale is increase with the epoch as predicted by $\Lambda$CDM. 

This evolution can be interpreted by the growth of structure in our universe. We show the accuracy of our method in the $5$ redshift bins for the NGC and SGC CMASS sample. And we show the $2\sigma$ consistency with $\Lambda$CDM.

\section{Conclusion $\&$ Discussion}
\paragraph{}In this study, we perform measurements of the homogeneity scale of our universe on DR12 BOSS CMASS galaxy sample. We measured the \textit{fractal correlation dimension} showing its scale dependence.$\mathcal{D}_{2}(r)$ reaches the homogeneous value asymptotically behaving as a homogeneous distribution on scales greater than $\mathcal{R}_{H} = 64.3\pm1.6h^{-1}Mpc$ at $z=0.538-0.592$. On scales less than the distribution behaves as a fractal for different scales.

Moreover, we have shown the consistency of our homogeneity scale measurement for different cuts of our data at the North and South galactic caps we have optimised the measurement of previous studies at $3 \%$ level.

Additionally, we present the epoch evolution of the cosmic homogeneity and its accordance with the $\Lambda$CDM prediction at percentage level.

Finally, since we assume the $\Lambda$CDM model to infer distances and to correct for redshift space distortions, we can only conclude with a consistency test validation of Cosmological Principle. 

\section*{Acknowledgments}

I am grateful for the organisers of the conference for inviting me to present the Homogeneity Scale Measurement of BOSS collaboration at the Moriond's meeting.


\label{Bibliography}


\bibliographystyle{plain} 

\bibliography{mybib}

\end{document}